\documentclass[english,usenatbib]{mn2e}
\usepackage[T1]{fontenc}
\usepackage[latin9]{inputenc}
\usepackage{xcolor}
\usepackage{amsmath}
\usepackage{graphicx}
\usepackage[authoryear]{natbib}
\PassOptionsToPackage{normalem}{ulem}
\usepackage{ulem}

\makeatletter

\providecolor{lyxadded}{rgb}{0,0,1}
\providecolor{lyxdeleted}{rgb}{1,0,0}

\DeclareRobustCommand{\lyxdeleted}[3]{{\color{lyxdeleted}\sout{#3}}}

\makeatother

\usepackage{babel}
\begin{document}

\title[Celestial Sphere Imaging]{Imaging on a Sphere with Interferometers: the Spherical Wave Harmonic
Transform}

\author[T. D. Carozzi]{T. D. Carozzi$^{1}$ \\%
$^{1}$Onsala Space Observatory, \\ Department of Earth and Space Sciences, \\ Chalmers University, Sweden}

\date{23 March 2015. This is a pre-copyedited, author-produced PDF of an
article accepted for publication in MNRAS following peer review.}
\maketitle
\begin{abstract}
I present an exact and explicit solution to the scalar (Stokes flux
intensity) radio interferometer imaging equation on a spherical surface
which is valid also for non-coplanar interferometer configurations.
This imaging equation is comparable to $w$-term imaging algorithms,
but by using a spherical rather than a Cartesian formulation this
term has no special significance. The solution presented also allows
direct identification of the scalar (spin 0 weighted) spherical harmonics
on the sky. The method should be of interest for future multi-spacecraft
interferometers, wide-field imaging with non-coplanar arrays, and
CMB spherical harmonic measurements using interferometers. 
\end{abstract}

\section{Introduction}

Basic radio interferometry deals with narrow fields-of-view measured
by antenna elements constrained to a plane. Under such conditions,
i.e. planar brightness distribution and planar visibility domain,
the van Cittert-Zernike (vCZ) theorem \citep{Thompson01} states that
the brightness and visibility distributions are two-dimensional Cartesian
Fourier transforms of each other. An extension of the van Cittert-Zernike
to arbitrarily wide fields and non-coplanar arrays was given in \citet{Carozzi2009a}
where it was found that the simple Fourier transform relation no longer
holds. 

The generalized vCZ relation given in \citet{Carozzi2009a} is still
similar to the original planar vCZ in that the brightness and visibility
domains are ultimately expressed in Cartesian coordinates. A different
approach to an interferometric relation for the full celestial sphere
was given in \citet{Macphie1975} which used spherical harmonics in
the visibility domain. However the main result in that paper was a
formula for point sources, i.e. the brightness distribution was given
in terms of delta functions on the sphere. More recently, \citet{McEwen2008}
used a spherical harmonic decomposition of visibility data to obtain
the celestial sky multipole moments, but their treatment of the radial
component of the visibility data was not made explicit.

In what follows I will provide a simple relation, analogous to the
vCZ, between a brightness distribution on the celestial sphere and
its visibility distribution in an arbitrary domain --- possibly non-coplanar
and not necessarily spherical --- using a special case of the spherical
Fourier-Bessel transform rather than using a Cartesian Fourier based
transform. 

The vCZ on a sphere relation presented here has several practical
applications. Spherical harmonics of the sky temperature derived from
interferometers is of current interest \citep{Kim2007,Ng2001}, and
in the future there are plans for a multi-spacecraft interferometer
mission. Such an interferometer would observe the full celestial sphere
rather than a hemisphere, which limits Earth based interferometers.
The results are also of interest to observations with non-coplanar
arrays that currently must deal with the so-called $w$-term \citep{Cornwell1992},
which is a consequence of adapting the two-dimensional, Cartesian
Fourier transform to work with three-dimensional visibility data to
produce images of the celestial sphere. Although the imaging technique
presented here is naturally suited to extended sources and multipole
moments, also narrow field-of-view interferometers could benefit since,
for high dynamic range imaging, the trend is to image the entire hemisphere
anyways in order to handle leakage from beam side-lobes.

\section{A relation between sky brightness on celestial sphere and non-coplanar
visibilities}

I start with the scalar intensity component of the extended vCZ theorem,
i.e. a relation between visibility $\mathcal{V}$ and brightness $B$,
as given in \citet{Carozzi2009a}, valid on the celestial sphere,
which can be written as 
\begin{align}
\mathcal{V}_{I}(\mathbf{r},k) & =\int B_{I}(\Omega_{k})\exp\left(-\mathrm{i}\mathbf{k}\cdot\mathbf{r}\right)\mathrm{d}\Omega_{k}\label{eq:Carozzi2009sphereAng}
\end{align}
where $\mathbf{r}$ is the position vector in the visibility domain,
$\mathbf{k}$ is the wavevector and $\Omega_{k}=(\theta_{k},\phi_{k})$
are the angular components of $\mathbf{k}$ on the sphere. The subscripts
$\bullet_{k}$ are used to denote that the angles refer to the spherical
components of the wavevector. Since I will only be concerned with
measurements in vacuum, the wavenumber $k=|\mathbf{k}|$ is equal
the frequency used for the visibility measurements divided by the
speed of light, $\omega/c$. Note that in eq. \eqref{eq:Carozzi2009sphereAng}
the phase reference position is the origin%
\footnote{That is I have removed the phase reference position so the interferometer
is not phased up towards any particular direction.%
}. The subscript $\bullet_{I}$ in the equation above denotes the Stokes
$I$ component, i.e. the scalar flux density. In what follows I discard
the Stokes $I$ subscript as I will only deal with this component.

The expression \eqref{eq:Carozzi2009sphereAng} actually implies that
$\mathcal{V}$ fulfills the Helmholtz equation, also known as the
wave equation. This fact is not well appreciated in the radio interferometry
literature, so I present it here. Operating with the Laplace operator
on eq. \eqref{eq:Carozzi2009sphereAng}, one finds that
\begin{equation}
\boldsymbol{\nabla_{r}}^{2}\mathcal{V}+k^{2}\mathcal{V}=0
\end{equation}
which is the Helmholtz equation, or wave equation, in the visibility
domain. The Helmholtz equation has, besides Cartesian solutions, also
solutions in spherical coordinates, and this suggests that there should
be a vCZ relation in terms of eigenfunctions of the spherical wave
equation, which are equal to \citep{Jackson99}
\begin{equation}
j_{\ell}(kr)Y_{\ell m}(\theta,\phi),\quad\textrm{for }\ell=0,1,2,\ldots;\: m=-\ell,\ldots,\ell\label{eq:sphericalwaveEigenf}
\end{equation}
where I invoke the boundary condition that the visibility should be
finite at the origin. $Y_{\ell m}(\Omega)$ is the standard, orthonormal
spherical harmonic function with $\ell,m$ corresponding to the polar
and azimuthal quantal numbers%
\footnote{Since the spherical harmonic quantal numbers $\ell$ and $m$ could
be confused with the standard notation for the direction cosines in
Cartesian Fourier imaging, the latter will not be used in this paper. %
} respectively, and $j_{\ell}(kr)$ is the spherical Bessel function
of the first kind. I will call these eigenfunctions \emph{spherical
wave harmonics}.

To fully convert eq. \eqref{eq:Carozzi2009sphereAng} into the eigenfunctions
given in eq. \eqref{eq:sphericalwaveEigenf}, I proceed as follows.
I use the plane wave decomposition formula, see \citet{Jackson99},
\begin{equation}
\mathrm{e}^{-\mathrm{i}\mathbf{k}\cdot\mathbf{r}}=4\pi\sum_{\ell=0}^{\infty}\sum_{m=-\ell}^{\ell}(-\mathrm{i})^{\ell}j_{\ell}(kr)Y_{\ell m}(\theta_{r},\phi_{r})Y_{\ell m}^{\ast}(\theta_{k},\phi_{k})\label{eq:PlaneWaveDecomposition}
\end{equation}
 where the subscripts $\bullet_{r}$ denote the spherical coordinates
of the visibility position vector $\mathbf{r}$ and $r=|\mathbf{r}|$.
When this is inserted into eq. \eqref{eq:Carozzi2009sphereAng} it
gives
\begin{align}
\mathcal{V} & =\int B(\Omega_{k})\left(4\pi\sum_{\ell=0}^{\infty}\sum_{m=-\ell}^{\ell}(-\mathrm{i})^{\ell}j_{\ell}(kr)Y_{\ell m}(\theta_{r},\phi_{r})Y_{\ell m}^{\ast}(\Omega_{k})\right)\mathrm{d}\Omega_{k}\nonumber \\
 & =4\pi\sum_{\ell=0}^{\infty}\sum_{m=-\ell}^{\ell}(-\mathrm{i})^{\ell}j_{\ell}(kr)Y_{\ell m}(\theta_{r},\phi_{r})\int B(\Omega_{k})Y_{\ell m}^{\ast}(\Omega_{k})\mathrm{d}\Omega_{k}.\label{eq:vCZplanewavEx}
\end{align}
Then I expand the brightness distribution into spherical harmonics
\begin{equation}
B(\Omega_{k})=\sum_{\ell=0}^{\infty}\sum_{m=-\ell}^{\ell}b_{\ell m}Y_{\ell m}(\Omega_{k})\label{eq:ComputeImageFromSpHcoef}
\end{equation}
where $b_{lm}$ are the multipole moments of the sky. Inserting this
back into eq. \eqref{eq:vCZplanewavEx} one obtains
\begin{multline}
\mathcal{V}=4\pi\sum_{\ell=0}^{\infty}\sum_{m=-\ell}^{\ell}(-\mathrm{i})^{\ell}j_{\ell}(kr)Y_{\ell m}(\theta_{r},\phi_{r})\\
\times\int\left(\sum_{\ell=0}^{\infty}\sum_{m=-\ell}^{\ell}b_{\ell m}Y_{\ell m}(\Omega_{k})\right)Y_{\ell m}^{\ast}(\Omega_{k})\mathrm{d}\Omega_{k}\\
=4\pi\sum_{\ell=0}^{\infty}\sum_{m=-\ell}^{\ell}(-\mathrm{i})^{\ell}j_{\ell}(kr)Y_{\ell m}(\theta_{r},\phi_{r})b_{\ell m}\label{eq:cartVisEqSphBri}
\end{multline}
where I have used the orthogonality relation for the spherical harmonic
functions 
\begin{equation}
\int_{0}^{4\pi}Y_{\ell m}(\Omega)Y_{\ell'm'}^{\ast}(\Omega)\mathrm{d}\Omega=\delta_{\ell\ell'}\delta_{mm'}.
\end{equation}
Finally I expand the visibility distribution into the spherical wave
harmonics, eq. \eqref{eq:sphericalwaveEigenf}, with coefficients
$\tilde{v}_{\ell m}$
\begin{equation}
\mathcal{V}=\sum_{\ell=0}^{\infty}\sum_{m=-\ell}^{\ell}\tilde{v}_{\ell m}j_{\ell}(kr)Y_{\ell m}(\Omega_{r}).\label{eq:VisibilitySphDecomp}
\end{equation}
see \citet{Jackson99}. Inserting this into the left-hand side of
eq. \eqref{eq:cartVisEqSphBri}, one obtains
\begin{multline}
\sum_{\ell=0}^{\infty}\sum_{m=-\ell}^{\ell}\tilde{v}_{\ell m}j_{\ell}(kr)Y_{\ell m}(\Omega_{r})\\
=4\pi\sum_{\ell=0}^{\infty}\sum_{m=-\ell}^{\ell}(-\mathrm{i})^{\ell}b_{\ell m}j_{\ell}(kr)Y_{\ell m}(\theta_{r},\phi_{r}).
\end{multline}
From this equation, due to the orthonormality of the $Y_{\ell m}$
harmonics, one can identify that for any $(\ell,m)$, 
\begin{align}
\tilde{v}_{\ell m} & =4\pi(-\mathrm{i})^{\ell}b_{\ell m}.\label{eq:mainresult}
\end{align}

Eq. \eqref{eq:mainresult} is an important result, and shows that
there is a simple proportionality relation between the brightness
distribution, in terms of $b_{\ell m}$, and the visibility distribution,
in terms of $\tilde{v}_{\ell m}$, with no integration or sum over
any domain. The simplicity of this result is due to the fact that
the spherical harmonic components are eigenfunctions of the measurement
equation on the sphere \eqref{eq:Carozzi2009sphereAng} and that these
components automatically fulfill the Helmholtz dispersion relation
$k^{2}=\omega^{2}/c^{2}$. By contrast, the Cartesian Fourier transform
consists of plane wave solutions, i.e. point sources, which are not
eigenfunctions of the measurement equation on the sphere and do not
automatically fulfill the dispersion relation which leads to the additional
complexity of dealing with the $w$-term, i.e. the third and final
wavevector component in the plane wave solutions.

\citet{McEwen2008} derived an essentially similar relationship to
\eqref{eq:mainresult}, albeit not explicitly. However, they did not
provide an explicit scheme to derive the harmonic coefficients for
an arbitrary array. In fact they speculated that a stable scheme could
be developed, arguing that the presence of zeros of the spherical
Bessel functions with large $\ell$ would complicate the recovery
of the coefficients. I argue that the zeros of the spherical Bessel
function for some $\ell$ simply mean that that particular $\ell$
does not contribute to the harmonic coefficient at that point, but
e.g. the spherical Bessel functions with $\ell\pm1$ will. In the
next section I will show that the radial part of the visibility can
indeed be incorporated into the recovery of the spherical harmonics
of the sky, and later I will show that, at least for $\ell\leq96$,
it is possible to produce images comparable to those made with the
Cartesian Fourier transform.

\section{Computing the spherical wave harmonic coefficients of the visibility
distribution}

The result expressed in eq. \eqref{eq:mainresult} shows that spherical
harmonic components of the celestial sky at a some frequency $\omega$
are proportional to a spherical Fourier-Bessel decomposition of the
visibility distribution with the corresponding $k$. Although this
vCZ relation is superficially simpler than the Cartesian Fourier transform,
it still implies a comparable computational complexity since the $v_{\ell m}$
components need to be determined from the interferometric measurements. 

In practice, an interferometer consists of an array of a finite number
of antennas from which complex voltages are measured, and the visibilities
are the complex powers obtained by cross-correlating between all antenna
pairs. Thus $\mathcal{V}$ can only be sampled at a finite set of
$Q$ measurements at points with spherical coordinates which I denote
as $\left\{ r_{i},\theta_{i},\phi_{i}\right\} _{i=1}^{Q}$. Note that
from now on I will dispense with the $r$ subscripts for the spherical
angles in the visibility domain that had been used in the previous
section. Although there are no formal restrictions on the sampling
distribution for the estimating the vCZ relation in the preceding
section, certain distributions will be more advantageous than others.

A detailed discussion of the numerical implementation is outside the
scope of the present paper, but a direct (non-gridded) naive solution
for $\tilde{v}_{\ell m}$ can be derived as follows. Consider a visibility
dataset $\mathcal{V}_{i}(k_{0})$ measured in narrow band with center
frequency $\omega_{0}$ sampled at arbitrary positions. These can
be seen as a sum of delta functions in the visibility domain
\begin{multline}
\mathcal{V}(k,r,\theta,\phi)=\sum_{i=1}^{Q}\mathcal{V}_{i}(k_{0})\delta(\mathbf{r}-\mathbf{r}_{i})\delta(k-k_{0})\\
=\sum_{i=1}^{Q}\frac{\mathcal{V}_{i}(k_{0})}{r^{2}\sin\theta}\delta(r-r_{i})\delta(\theta-\theta_{i})\delta(\phi-\phi_{i})\delta(k-k_{0})\label{eq:deltasInVisibility}
\end{multline}
where $k_{0}=\omega_{0}/c$ and the factor in the denominator is the
normalization factor for the delta functions in spherical coordinates.
The delta function in the $k$ domain is a simplifying approximation
of the spectral density of the frequency band response function. Multiplying
the right-hand side of eq. \eqref{eq:deltasInVisibility} with $j_{\ell}(k_{0}r)Y_{\ell m}^{\ast}(\theta,\phi)$,
and then integrating this over\lyxdeleted{tobia carozzi,,,,}{Fri Apr 17 12:13:46 2015}{
} a\lyxdeleted{tobia carozzi,,,,}{Fri Apr 17 12:13:46 2015}{ } spherical
volume that bounds\lyxdeleted{tobia carozzi,,,,}{Fri Apr 17 12:13:46 2015}{
} the visibility domain results in
\begin{multline}
\int_{0}^{\infty}\!\!\int_{0}^{\pi}\!\!\int_{0}^{2\pi}\sum_{i=1}^{Q}\frac{\mathcal{V}_{i}(k)}{r^{2}\sin\theta}j_{\ell}(k_{0}r)Y_{\ell m}^{\ast}(\theta,\phi)\delta(r-r_{i})\delta(\theta-\theta_{i})\delta(\phi-\phi_{i})\\
\times\delta(k-k_{0})r^{2}\sin\theta\,\mathrm{d}r\mathrm{d}\theta\mathrm{d}\phi=\sum_{i=1}^{Q}\mathcal{V}_{i}(k_{0})j_{\ell}(k_{0}r_{i})Y_{\ell m}^{\ast}(\theta_{i},\phi_{i})\delta(k-k_{0}).
\end{multline}
For the left-hand side of eq. \eqref{eq:deltasInVisibility}, I insert
eq. \eqref{eq:VisibilitySphDecomp} and do exactly the same the steps
as were performed on the right-hand side and then get
\begin{multline}
\int_{0}^{\infty}\!\!\int_{0}^{\pi}\!\!\int_{0}^{2\pi}\sum_{\ell'=0}^{\infty}\sum_{m'=-\ell'}^{\ell'}\tilde{v}_{\ell'm'}j_{\ell'}(kr)Y_{\ell'm'}(\theta,\phi)j_{\ell}(k_{0}r)Y_{\ell m}^{\ast}(\theta,\phi)\\
\times r^{2}\sin\theta\,\mathrm{d}r\mathrm{d}\theta\mathrm{d}\phi=\sum_{\ell=0}^{\infty}\sum_{m=-\ell}^{\ell}\tilde{v}_{\ell m}\intop_{0}^{\infty}j_{\ell}(kr)j_{\ell}(k_{0}r)r^{2}\,\mathrm{d}r\\
=\sum_{\ell=0}^{\infty}\sum_{m=-\ell}^{\ell}\tilde{v}_{\ell m}\frac{\pi\delta(k-k_{0})}{2k_{0}^{2}}\label{eq:lefthandsideDeltasEq}
\end{multline}
where I have used the relation
\begin{equation}
\int_{0}^{\infty}j_{\ell}(kr)j_{\ell}(k_{0}r)r^{2}\,\mathrm{d}r=\frac{\pi\delta(k-k_{0})}{2k_{0}^{2}}\label{eq:righthandsideDeltasEq}
\end{equation}
which is valid for all $\ell$, see \citet{Leistedt2012}. Integrating
both the left- and right-hand sides, i.e. the last results of eq.
\eqref{eq:lefthandsideDeltasEq} and eq. \eqref{eq:righthandsideDeltasEq},
over all $k$ and equating these two results I find that 
\begin{equation}
\tilde{v}_{\ell m}(k_{0})=\frac{2k_{0}^{2}}{\pi}\sum_{i=1}^{Q}\mathcal{V}_{i}(k_{0})j_{\ell}(k_{0}r_{i})Y_{\ell m}^{\ast}(\theta_{i},\phi_{i}).\label{eq:ExplicitVisDecompFormal}
\end{equation}
This my main result in terms of providing an explicit, direct quadrature
rule for computing the spherical wave coefficients from arbitrarily
placed visibility samples. Note that there is no formal restriction
on the radial positions of the samples, for instance with respect
to the zeros of the spherical Bessel functions.

The transform used above to derive eq. \eqref{eq:ExplicitVisDecompFormal}
is a type of spherical Bessel-Fourier decomposition, see \citet{Leistedt2012,Baddour2010}.
But a crucial difference is that, in the present work, the radial
component of the wavevector, i.e. the wavenumber $k$, is already
known since radio interferometric visibility data is almost always
given as functions of frequency in narrow bands, hence the delta function
in $k$. Thus the transform is two-dimensional rather than three-dimensional
for a given frequency. For this reason it may be more appropriate
to call this something else instead, so I will use the term\lyxdeleted{tobia carozzi,,,,}{Fri Apr 17 12:13:46 2015}{
}\emph{ spherical wave harmonic transform} (SWHT).

\section{All-sky imaging examples}

In this section I apply the SWHT to real radio interferometer data
to illustrate the imaging technique. I used data from the Swedish
LOFAR station, known as SE607, in particular the Low Band Array (LBA),
which is a \textasciitilde{}60 m diameter array of 96 crossed dipoles
placed in a pseudo-random, co-planar pattern covering the frequency
range $10-90$ MHz.

The dataset I used was a snap-shot, i.e. the cross-correlations (and
auto-correlations) are integrated over a short, 10 s, time interval,
so that the array can be taken to be co-planar. This was chosen since
it then can be compared with the ordinary (non-gridded) Cartesian
Fourier transform. The data was for center frequency 37.1 MHz in a
192 kHz wide band. I applied eq. \eqref{eq:ExplicitVisDecompFormal}
to the visibility data and computed the $\tilde{v}_{\ell m}$ coefficients
up to $\ell=L_{\textrm{max}}=96$, which matches the number of elements.
These coefficients were then converted to the sky harmonic coefficients
$b_{\ell m}$ using eq. \eqref{eq:mainresult}, and then these coefficients
were used to generate an image through eq. \eqref{eq:ComputeImageFromSpHcoef}. 

\begin{figure*}
\noindent \begin{centering}
\includegraphics[width=1\textwidth]{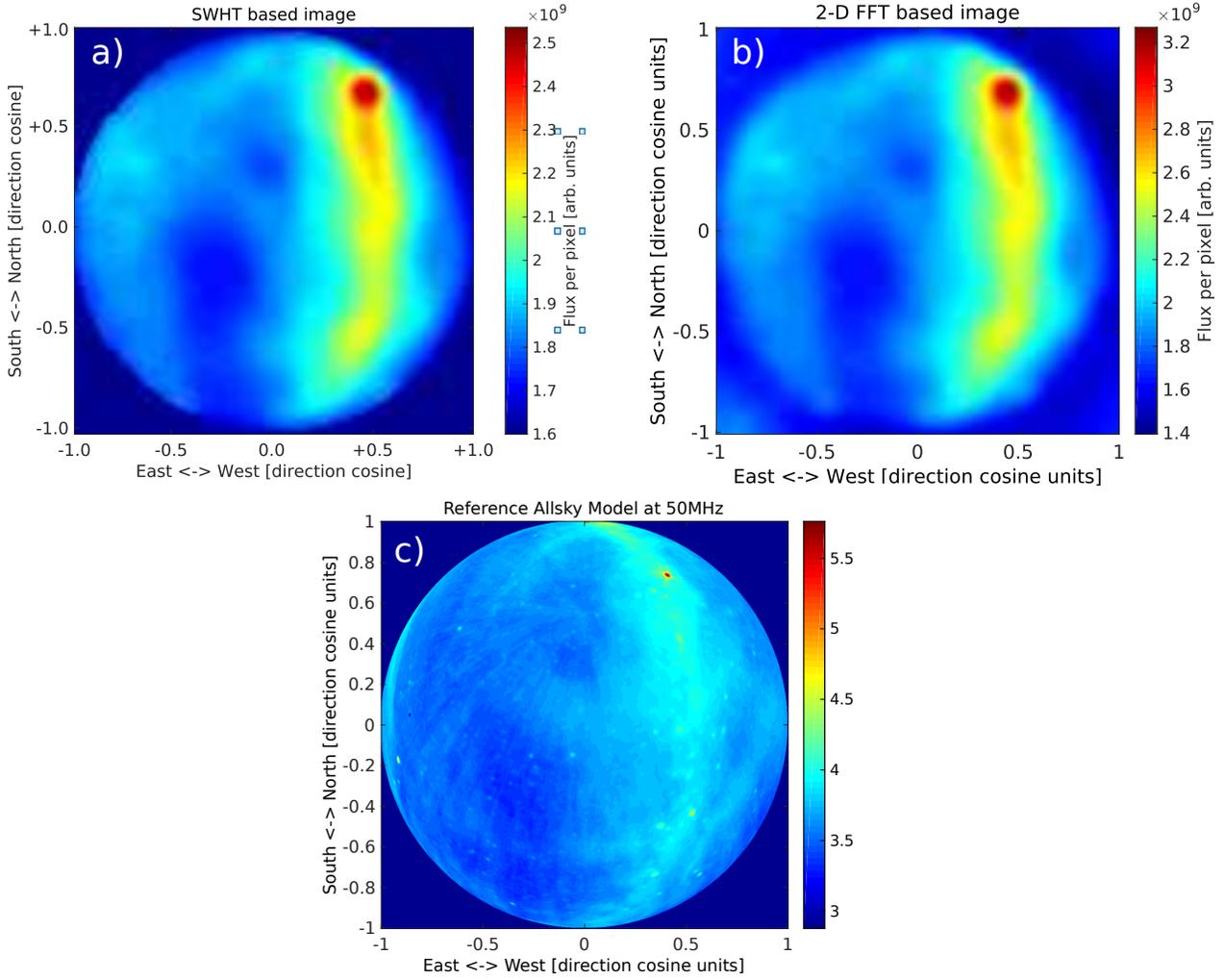}
\par\end{centering}

\caption{Orthographically projected images of Stokes $I$ flux in arbitrary
units of the celestial hemisphere over LOFAR SE607 LBA on 3 December
2014, 00:16 UT. The subplot a) image was computed using the spherical
Bessel harmonics with $\max(\ell)=96$ at 37.1 MHz. The subplot b)
image was computed using a non-gridded two-dimensional Fourier transform
also at 37.1 MHz. Subplot c) is an reference model image at 50 MHz.
The extended emission in the West is the Milky Way. The strong point
source in the North-West is Cassiopeia A. Emission from the galactic
North spur can be seen in the North-East. The circle suggested in
these images is the telescope local horizon at elevation 0. Note that
neither of the images a) or b) have been compensated for the antenna
gain pattern. \label{fig:imageTechnCompar} }
\end{figure*}

The result of this SWHT technique is shown in Figure \ref{fig:imageTechnCompar},
subplot a). For comparison, subplot b) shows the ordinary Cartesian
Fourier transformed image, also known as a dirty image, and subplot
c) shows a reference model at the slightly higher frequency of 50
MHz and with better resolution. It is clear from this Figure that
the SWHT is very similar to the Cartesian Fourier transform. The main
difference is the presence of emissions beyond the telescope horizon,
i.e. directions apparently below 0 elevation, for instance in the
South-East corner of subplot b). These emissions are an erroneous
artifact of the two-dimensional Cartesian Fourier imaging technique,
since the two Cartesian Fourier components, or direction cosines,
go from -1 to +1, and there is nothing to stop components with absolute
value greater than 1 from contributing to the image. In other words,
this illustrates the fact that the Cartesian Fourier transform does
not automatically fulfill the dispersion relation $k^{2}=\omega^{2}/c^{2}$,
as already mentioned.

The run times were slower for the SWHT, but these could be improved
up by using fast spherical harmonic transform algorithms, such as
\citet{Rokhlin2006}, which have computation time complexity of the
order $\mathcal{O}(N^{2}\log N)$, where $N$ is the number of sample
points. It is possible that an SWHT algorithm could be constructed
to have a time complexity not much greater than this, making it comparable
to the $w$-term imaging algorithms.

\section{Conclusions}

I have derived a vCZ relation, eq. \eqref{eq:mainresult}, between
a spherical
brightness distribution and an unconstrained visibility distribution.
I have also presented the SWH transform of the visibility data to
compute the spherical harmonics of the sky from which images can be
made. This technique was shown to be capable of producing images comparable
to ordinary dirty images. It should be useful for radio inferometric
imaging of extended sources or for determining multipole moments of
the celestial sky. It extends naturally to visibility data from non-coplanar
arrays, and thus the technique is comparable to $w$-term imaging
methods.


\begin{thebibliography}{11}
\providecommand{\natexlab}[1]{#1}
\providecommand{\url}[1]{\texttt{#1}}
\expandafter\ifx\csname urlstyle\endcsname\relax
  \providecommand{\doi}[1]{doi: #1}\else
  \providecommand{\doi}{doi: \begingroup \urlstyle{rm}\Url}\fi

\bibitem[Baddour(2010)]{Baddour2010}
Natalie Baddour.
\newblock Operational and convolution properties of three-dimensional fourier
  transforms in spherical polar coordinates.
\newblock \emph{JOSA A}, 27\penalty0 (10):\penalty0 2144--2155, 2010.

\bibitem[Carozzi and Woan(2009)]{Carozzi2009a}
T.~D. Carozzi and G.~Woan.
\newblock A generalized measurement equation and van {C}ittert-{Z}ernike
  theorem for wide-field radio astronomical interferometry.
\newblock \emph{Monthly Notices of the Royal Astronomical Society},
  395\penalty0 (3):\penalty0 1558--1568, May 2009.
\newblock \doi{10.1111/j.1365-2966.2009.14642.x}.
\newblock URL \url{http://dx.doi.org/10.1111/j.1365-2966.2009.14642.x}.

\bibitem[{Cornwell} and {Perley}(1992)]{Cornwell1992}
T.~J. {Cornwell} and R.~A. {Perley}.
\newblock {Radio-interferometric imaging of very large fields - The problem of
  non-coplanar arrays}.
\newblock \emph{Astronomy and Astrophysics}, 261:\penalty0 353--364, July 1992.

\bibitem[Jackson(1999)]{Jackson99}
John~D. Jackson.
\newblock \emph{Classical Electrodynamics}.
\newblock Wiley~\&~Sons,~Inc., New~York,~NY~\ldots, third edition, 1999.
\newblock ISBN~0-471-30932-X.

\bibitem[Kim(2007)]{Kim2007}
Jaiseung Kim.
\newblock Direct reconstruction of spherical harmonics from interferometer
  observations of the cosmic microwave background polarization.
\newblock \emph{Monthly Notices of the Royal Astronomical Society},
  375\penalty0 (2):\penalty0 625--632, 2007.
\newblock \doi{10.1111/j.1365-2966.2006.11285.x}.
\newblock URL \url{http://mnras.oxfordjournals.org/content/375/2/625.abstract}.

\bibitem[Leistedt et~al.(2012)Leistedt, Rassat, Réfrégier, and
  Starck]{Leistedt2012}
B.~Leistedt, A.~Rassat, A.~Réfrégier, and J.-L. Starck.
\newblock 3dex: a code for fast spherical fourier-bessel decomposition of 3d
  surveys.
\newblock \emph{A\&A}, 540:\penalty0 A60, 2012.
\newblock \doi{10.1051/0004-6361/201118463}.
\newblock URL \url{http://dx.doi.org/10.1051/0004-6361/201118463}.

\bibitem[Macphie and Okongwu(1975)]{Macphie1975}
Robert~H. Macphie and E.~H. Okongwu.
\newblock Spherical harmonics and earth-rotation synthesis in radio astronomy.
\newblock \emph{Antennas and Propagation, IEEE Transactions on}, 23\penalty0
  (3):\penalty0 386--391, 1975.

\bibitem[McEwen and Scaife(2008)]{McEwen2008}
J.~D. McEwen and A.~M.~M. Scaife.
\newblock Simulating full-sky interferometric observations.
\newblock \emph{Monthly Notices of the Royal Astronomical Society},
  389\penalty0 (3):\penalty0 1163--1178, 2008.
\newblock \doi{10.1111/j.1365-2966.2008.13690.x}.
\newblock URL
  \url{http://mnras.oxfordjournals.org/content/389/3/1163.abstract}.

\bibitem[Ng(2001)]{Ng2001}
Kin-Wang Ng.
\newblock Complex visibilities of cosmic microwave background anisotropies.
\newblock \emph{Phys. Rev. D}, 63:\penalty0 123001, May 2001.
\newblock \doi{10.1103/PhysRevD.63.123001}.
\newblock URL \url{http://link.aps.org/doi/10.1103/PhysRevD.63.123001}.

\bibitem[Rokhlin and Tygert(2006)]{Rokhlin2006}
Vladimir Rokhlin and Mark Tygert.
\newblock Fast algorithms for spherical harmonic expansions.
\newblock \emph{SIAM Journal on Scientific Computing}, 27\penalty0
  (6):\penalty0 1903--1928, 2006.

\bibitem[Thompson et~al.(2001)Thompson, Moran, and Swenson]{Thompson01}
A.~Richard Thompson, M.~Moran Moran, and George W.~Jr Swenson.
\newblock \emph{Interferometry and {S}ynthesis in {R}adio {A}stronomy}.
\newblock John Wiley \& Sons, Inc., 2001.

\end{thebibliography}
\end{document}